\documentclass[a4paper]{raa}            
\usepackage{graphicx,times}             
\usepackage[numbers]{natbib}            
\usepackage[dvips,colorlinks=true, linkcolor=red,urlcolor=magenta,citecolor=blue,anchorcolor=green]{hyperref}
\usepackage{color}
\def\vp{\color{blue}}
\bibpunct[,]{[}{]}{,}{n}{}{,}       

\usepackage[figuresright]{rotating}      
\usepackage{lscape,longtable}
\usepackage{enumerate}
\begin{document}

   \title{Hybrid Pulsators --- Pulsating Stars with Multiple Identities
\,$^*$
\footnotetext{$*$ Web version at \url{http://www.chjaa.org/COB/hybrid/}.}
}

   \volnopage{Vol.15 (2015) No.6, 000--000}      
   \setcounter{page}{1}          

   \author{A.-Y. Zhou
   }

   \institute{National Astronomical Observatories, Chinese Academy of Sciences,
             Beijing 100012, China; {\it aiying@nao.cas.cn}\\
   }

   \date{Received~~2015 month day; accepted~~2015~~month day}

\abstract{ We have carried out a statistic survey on the pulsating variable stars with
multiple identities. These stars were identified to exhibit two types of pulsation
or multiple light variability types in the literature, and are usually called hybrid pulsators.
We extracted the hybrid information based on the Simbad database.
Actually, all the variables with multiple identities are retrieved.
The survey covers various pulsating stars across the Hertzsprung-Russell diagram.
We aim at giving a clue in selecting interesting targets for further observation.
Hybrid pulsators are excellent targets for asteroseismology.
An important implication of such stars is their potential in advancing the theories
of both stellar evolution and pulsation.
By presenting the statistics,
we address the open questions and prospects regarding current status of hybrid pulsation studies.
}

   \authorrunning{A.-Y. Zhou}            
   \titlerunning{Hybrid Pulsating Stars }  

   \maketitle

%
%
\section{Motivation}           
\label{sect:intro}
\begin{figure}[t!!!]
   \vspace{2mm}
\hspace{-6mm}   \includegraphics[width=146mm, height=110mm,angle=0,scale=1.1]{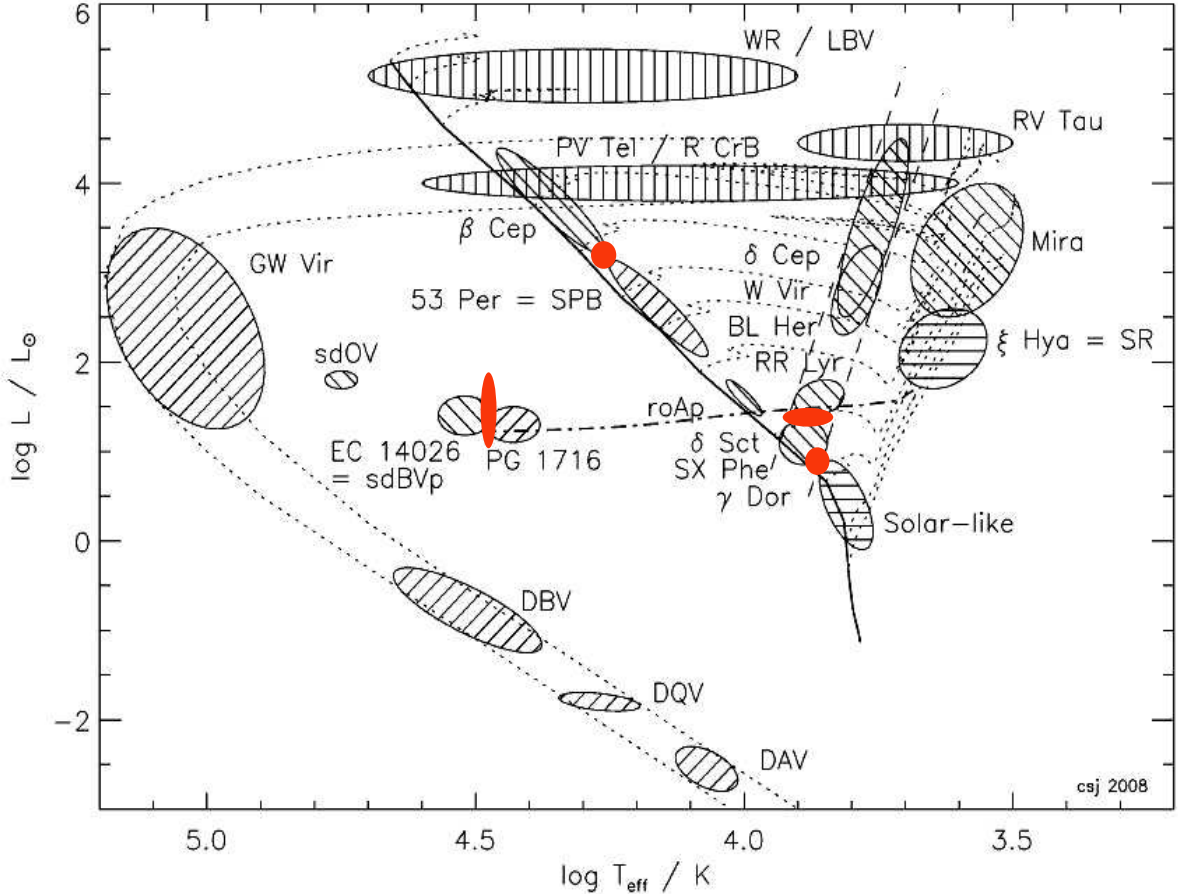}
\begin{minipage}[]{159mm}
   \caption{A version of the Hertzsprung-Russell diagram of pulsating stars with
   four indicative overlapping domains.
   Adopted from Jeffery (2008).}\end{minipage}
   \label{Fig:HR-diagram}
\end{figure}

Pulsating stars are a kind of intrinsic variable stars, which show periodic brightness/luminosity fluctuations
due to gradient opacities on the outer layers and in deep interior of the stars.
Different types of pulsating stars are distinguished by their periods of pulsation and the shapes of their light curves.
They are located in different instability strips on the Hertzsprung-Russell diagram (see e.g. Fig~\ref{Fig:HR-diagram}).
The stars in different areas have distinctly different physical properties and
are in different evolutionary stages of their life.
These instability domains are not necessarily distinct, but may be overlapped.
Stars having two different sets of pulsation spectra excited simultaneously may therefore exist
within overlapping instability strips.
When a pulsating variable star exhibits two types of pulsation,
or it is identified to be multiple variability types,
we usually call it `a hybrid pulsator'.

Stellar pulsation (frequency) carries valuable information on the physical condition within the star.
According to pulsational theory, oscillations must be trapped in some part of the stellar interior,
i.e. different mode types of pulsation are trapped in different regions in the star's interior,
and are driven by different excitation mechanisms.
Low-order non-radial modes have mixed character: acoustic-type in the envelope and
gravity-type in the deep radiative interior.
Three types of oscillation have been detected widely in stars:
(1) the stochastic oscillations generated by convection near stellar surface are observed in the Sun and solar-like stars;
(2) pressure modes ($p$ modes) refer to the outer convection layers of a star;
while (3) gravity modes ($g$ modes) penetrate deep into the radiative zone.
Hybrid pulsators having two mode types such as $p$- and $g$-modes simultaneously are of
particular interesting since the two mode types probe different regions within the star.
A hybrid pulsator is thus much interesting than those pulsators of single types.
Hybrid pulsators have excellent asteroseismic potential, as multiple types of oscillation can be
exploited to obtain a more complete picture of the stellar interior (Handler 2009).
A significant implication of studying hybrid pulsators is the answer why these stars present
mixed pulsations. One the other hand, hybrid pulsators may lead us to better understand
the undergoing physical processes of stellar interior and the evolution of the stars.

We could expect that hybrid pulsators would exist in the conjunction regions of the
pulsating variable stars distributed in the H-R diagram.
Is the occurrence of hybrids in the H-R diagram confined to the overlapping region of
two neighbouring instability domains?
On the contrary, can we see an occurrence of hybrid pulsations presented in
two non-adjacent classes of pulsators in the H-R diagram?
In recent years, a number of hybrid pulsators have been revealed.
We take three representative kinds of hybrid pulsators
for examples (see Fig.~\ref{Fig:HR-diagram} and Table~\ref{Tab:hybrid-samples}).

Instability domains of $\delta$ Scuti and $\gamma$ Doradus stars overlap in the H-R diagram,
the existence of hybrid $\delta$ Sct -- $\gamma$ Dor pulsators may appear evident,
and it was truly predicted by Dupret et al. (2004) and
the first unambiguous detection was reported by Henry \& Fekel (2005).
V529 And (=HD 8801, the first single Am star pulsating with $p$ and $g$ modes),
HD 49434 (Brunsden et al. 2014), and TYC 3139-151-1 are three of the confirmed
$\delta$ Sct -- $\gamma$ Dor hybrid pulsators.
That means we detected both pressure- and gravity-caused non-radial pulsation in them.
Moreover, $\gamma$ Dor variability was also detected in a rapidly oscillating [peculiar] A-type star (roAp),
and solar-like oscillations were discovered in a $\delta$ Sct star.
Handler (2012) presented a review on the hybrid pulsators among A/F-type stars.
KIC 3858884 is a hybrid $\delta$ Sct pulsator in a highly eccentric eclipsing binary (Maceroni et al. 2014),
in which the pulsating second component showed high-order $g$-modes apart from
the typical $\delta$ Sct pulsation frequencies of the primary component.
Such a system of eclipsing binaries containing non-radial pulsators allows
combining two different and independent sources of information on the internal structure and
evolutionary status of the components and, at the same time,
studying the effects of tidal forces on pulsations.
There are many hybrid $\delta$ Sct--$\gamma$ Dor pulsators (see a recent investigation
by Hareter 2013).

Six slowly pulsating B stars (SPB, main-sequence stars of spectral types B2 to B9)
had been found to exhibit both SPB (high-order gravity modes) and $\beta$ Cephei-type
(low-order acoustic and gravity modes) variability:\\
$\nu$ Eridani (=HD 29248, Handler et al. 2004, Dziembowski \& Pamyatnykh 2008, Daszy\'{n}ska-Daszkiewicz \& Walczak 2010),
$\iota$ Herculis (=HR 6588=HD 160762, BCEP),
$\gamma$ Pegasi (also a binary system, Chapellier et al. 2006, Pandey et al. 2011),
53 Arietis,
V354 Persei(=HD 13745),
53 Piscium (Le Contel et al. 2001),
and 12 Lacertae (Handler et al. 2006, Dziembowski \& Pamyatnykh 2008).
As a result of our survey, there are 20 $\beta$ Cep--SPB hybrid pulsators.
It is interesting to examine whether the presentation of both high-order $g$ modes and
low-order $p$ and $g$ modes property is consistent with the theoretical expectations.

Hybrid pulsating subdwarf B stars (sdBV) have been revealed. In fact, these hybrid sdBV pulsators showing
both pressure and gravity modes form a subclass of sdBV, i.e. sdBV$_{\rm rs}$. For instance,
the prototype, HS 0702+6043 (=DW Lyn, Schuh et al. 2006, Lutz et al. 2008),
Balloon 090100001 (Baran et al. 2005),
HS 2201+2610 (=V391 Peg, Lutz et al. 2008).
In addition to the hybrid pulsational behaviour, HS 2201+2610 is also a pulsator
hosting an orbiting planet (Silvotti et al. 2007, Lutz et al. 2008).
J08069+1527 is a newly discovered high-amplitude, hybrid subdwarf B pulsator (Baran et al. 2011).

The goal of this survey is to have an overall view of various hybrid pulsators including
the three classes mentioned above. In fact, almost all stars with multiple object types
are surveyed, such as stars like HD 25558, a long-period double-lined binary with
two SPB components (S\'{o}dor et al. 2014), other pulsating stars in binary systems, and so on.
We first extracted the information based on the Simbad database.
This report presents our preliminary results, the cross table Table~\ref{Tab:hybridPul}.
In the next step, we plan to update the table by checking literature and other database.
In the mean time, we are going to do observations on some of the interesting hybrid pulsators,
especially the pulsators hosting a transiting exoplanet.

\section{Statistics of Hybrid Pulsating Variable Stars}
As Simbad provides the largest cross-identifications of all kinds of celestial objects (up to 7,709,060 by 2014 December 30),
covering various surveys and published catalogs, we extract the objects with
multiple identities by setting a criteria of dual object types.
We must point out that the presented results in the cross table Table~\ref{Tab:hybridPul} is
based solely on the Simbad database.
Although Simbad archives data from 13187+ published catalogs, it certainly has a lag in
updating data because it does not reflect the latest status of publications
in terms of a simple literature search on ADS, as demonstrated in Table~\ref{Tab:hybrid-samples}.

Moreover, Simbad does not check the reality of a star's multiple identities.
Therefore, bias and even misclassification can be seen. For example,
the $\delta$ Scuti star V577 Oph was classified as RR Lyr-type;
the roAp prototype HD 101065(=V816 Cen) was not grouped in roAp stars but misclassified
as a $\delta$ Sct star;
the SPB prototype 53 Persei (=V469 Per) was misclassified as
$\beta$ Cep type (indicating mixture or confusion between $\beta$ Cep and SPB sometimes),
the SPB star FR Sct was misclassified to be an symbiotic star;
the semi-regular prototype $\xi$ Hya was classified as a non-variable, high proper-motion star,
rather than its true type; and so on.
Obviously, there are objects with confusion classifications.
On the other hand, roAp stars, SPB, sdBV, $\alpha$ Cyg-type pulsators, and others
were not explicitly classified, instead Simbad gathered them in the group `[pulsating] variable stars'.
For instance, the long-period (slow) $g$-mode sdBVs PG 1716+426 (=V1093 Her) was listed to be
a spectroscopic binary as main object type together with the white dwarf type.
So we are unable to extract directly the stars belonging to several specified classes,
including sdBV\footnote{Actually we retrieved the Simbad by spectral type 'sdB' together with object type 'V*' for sdBV.},
SPB, $\alpha$ Cyg-type pulsating stars, etc.

\begin{table}[t!]
\caption[c]{Samples of Representative Hybrid Pulsators.}
   \label{Tab:hybrid-samples}
   \vspace{-5mm}
   \begin{center}
   \begin{tabular}{lll}
\hline\noalign{\smallskip}
Families of Hybrids     & Hybrid Pulsators      & References    \\
   \hline\noalign{\smallskip}
$\beta$ Cep--SPB hybrids          & $\nu$ Eridani = HD 29248& Handler et al.(2004), Daszy\'{n}ska-Daszkiewicz \& Walczak(2010) \\
$\beta$ Cep--SPB hybrids          & $\gamma$ Pegasi   & a binary system, Chapellier et al.(2006), Pandey et al.(2011)\\
$\beta$ Cep--SPB hybrids          & $\iota$ Herculis  &=HR 6588=HD 160762, BCEP, \\
$\beta$ Cep--SPB hybrids          & 53 Piscium        & Le Contel et al. (2001)            \\
$\beta$ Cep--SPB hybrids          & 12 Lac            & Handler et al. (2006), Dziembowski \& Pamyatnykh (2008)\\
%
hybrid $\delta$ Sct--$\gamma$ Dor & V529 And          &=HD 8801, 1 of the two identified by Simbad, Handler (2009)\\
hybrid $\delta$ Sct--$\gamma$ Dor & TYC 3139-151-1    &=KIC 4749989, 1 of the two identified by Simbad\\
hybrid $\delta$ Sct--$\gamma$ Dor & BD+18 4914        & by MOST, Rowe et al. (2006)\\
hybrid $\delta$ Sct--$\gamma$ Dor & HD 114839         & by MOST, King et al. (2006), Hareter et al. (2011)\\
hybrid $\delta$ Sct--$\gamma$ Dor & CoRoT 102699796   & first metal-poor Herbig Ae pulsator (Ripepi et al. 2011) \\
hybrid $\delta$ Sct--$\gamma$ Dor & CoRoT 105733033   & Chapellier et al. 2012\\
hybrid $\delta$ Sct--$\gamma$ Dor & CoRoT 100866999   & in EB, Chapellier \& Mathias (2013)\\
hybrid $\delta$ Sct--$\gamma$ Dor & HD 49434          & Brunsden et al. (2014)\\
hybrid $\delta$ Sct--$\gamma$ Dor & KIC 6761539       & Herzberg et al. (2012)\\
hybrid $\delta$ Sct--$\gamma$ Dor & KIC 8569819       & in EA, Kurtz et al. (2015)\\
hybrid sdBV: sdBV$_{\rm rs}$      & HS0702+6043       &=DW Lyn, the prototype, Schuh et al.(2006), Lutz et al.(2008)\\
hybrid sdBV: sdBV$_{\rm rs}$      & Balloon 090100001 & Baran et al. (2005)\\
hybrid sdBV: sdBV$_{\rm rs}$      & HS 2201+2610      &=V391 Peg, planet-hosting sdBV, Lutz et al.(2008)\\
hybrid sdBV: sdBV$_{\rm rs}$      & RAT0455+1305      & Baran \& Fox-Machado (2009)\\
hybrid sdBV: sdBV$_{\rm rs}$      & J08069+1527       & Baran et al. (2011)\\
hybrid PG 1159                    & HS 2324+3944      & =V409 And, DOV, pulsating pre-white dwarf, Handler et al.(1997)\\
hybrid PG 1159                    & Abell 43          & Vauclair et al. (2005)\\
hybrid B-type pulsator            & HD 25558          & =V1133 Tau, double-lined binary of SPB+SPB components\\
   \hline\noalign{\smallskip}
   \end{tabular}
   \end{center}
\end{table}

Simbad currently resolved only two $\delta$ Sct--$\gamma$ Dor hybrid pulsators (V529 And, TYC 3139-151-1),
but actually more than ten such stars are identified in the literature (see Table~\ref{Tab:hybrid-samples}).
One can make a hypothesis that all objects in the overlapping region between $p$- and $g$-modes
are hybrid stars. In their close vicinity there are two other stars with only $p$-modes
whileas a bunch of stars with only $g$-modes detected (refers to fig.8 of Baran \& Fox-Machado 2009).
Based on the CoRoT LRa01 light curves, there are 418 $\gamma$ Dor candidates,
274 $\delta$ Sct--$\gamma$ Dor hybrid candidates were resolved (see p.18 of Hareter 2013).
While the Kepler mission reports a large sample of $\gamma$ Dor stars (Tkachenko et al. 2013).

As such, the numbers of each kind of pulsating stars as well as the statistics of hybrid pulsational behaviour
are not by no means indisputable.
Missing of member stars in a class of pulsators as well as discrepancies between the Simbad data and
other existing catalogs or database (e.g, ADS, CoRoT, Kepler data archives, etc.) are common.
Taking the above into consideration, one needs to re-check
the Simbad and other database for an update before using the data.
In this regard, a web-based version of the present cross table (Table~\ref{Tab:hybridPul})
has been developed at
\url{http://www.chjaa.org/COB/hybrid/}
\footnote{The web page in fact provides a friendly accessing interface to
the Simbad database in such a way that one can enquiry any kind of hybrids
among various variable stars. The output format of the resulting page was initialized
through a script acting the similar function of the Simbad's `Output options' settings.
So one can re-configure the `Output options' to view more parameters.},
which aims at providing the latest status of hybrid variability types.
Future updates will be first made available on the web.

In addition, we performed a relevant statistics on the major types of variable stars.
As of 2014 December 31, Simbad has archived 7709060 objects --- this number changes daily,
of which 3754965 and 248103 are stars and variable stars, respectively,
the rest are non-stellar objects (galaxies, planetary nebulae, etc.).
GCVS4.2 collected 58947 designated known variables (including 10979 extragalactic variables, Samus et al. 2012),
which accounts for only 23.7\% of the Simbad's variables. The detailed percentages of
each class of variables in the whole database are given in Table~\ref{Tab:percent}.
Some interesting hybrid pulsators are collected in Table~\ref{Tab:demo-hybrid},
where the reality of multiple identities was not examined one by one.
Because the leading columns of Tables~\ref{Tab:hybridPul}--\ref{Tab:demo-hybrid}
are the Simbad's acronyms, which are not widely used by astronomers,
a lookup table (Table~\ref{Tab:object-type}) for the acronyms is provided.
This table also lists the numbers of stars in each class.

In both theoretical and observational points of view, we would draw the readers' attention towards
the followings:
\begin{enumerate}[(1)]
  \item Pay close attention to the zero occurrences of any combinations of different types
  of pulsators in Table~\ref{Tab:hybridPul},
  especially those overlapped or adjacent regions on the H-R diagram

  \item There are actually multiple identities for the hybrid pulsators, check the details for any one hybrid.

  \item Close attention should be also paid to the binary systems containing pulsating stars.
  There are quite a number of EB, SB, ELL, and multiple-star systems containing pulsating component[s].
  We refer readers to \href{http://arxiv.org/abs/1002.2729}{\vp Zhou (2010)} for further information.

  \item Observers are encouraged to look at the 'EP' entry --- the planetary transit systems.
  More efforts are deserved to searching for various possible pulsators containing exoplanets.
  By now, at least two sdBV stars are known to have planets:
  V391 Pegasi was the first known sdB planet-host, and
  Kepler-70 has a system of close-orbiting planets that may be the remnants of a giant planet
  that was engulfed by the red giant progenitor.
\end{enumerate}

\section{Astrophysical Implications and Open Questions}

Apart from enhancing observational efforts in hybrid pulsators, we address the theoretic studies
on the excitation of different types of pulsational modes in hybrid pulsators.
Several attempts are found for reference:
\begin{enumerate}[(i)]
  \item  Can opacity changes help to reproduce the hybrid star pulsations in $\beta$ Cep--SPB
  dual pulsators (Zdravkov \& Pamyatnykh 2008)? From a theoretical point of view,
  Pamyatnykh (1999) and Miglio et al.(2007) showed that with OP opacities (Seaton 2005),
  the instability strips of $\beta$ Cep and SPBs intersect, so that theoretical models with
  both unstable $p$-modes and high order $g$-modes exist.

  \item  Improvements in current models for hybrid sdBV, e.g. the driving mechanism for
  long-period, high-order $g$-modes in long-period pulsating subdwarf B stars, given by
  Fontaine et al.(2003).

  \item  Driving and damping mechanisms in hybrid pressure--gravity modes pulsators (Dupret et al. 2009):
$\gamma$ Dor instability strip overlaps the red edge of the $\delta$ Sct.
$\gamma$ Dor are stars pulsating with high order $g$-modes, apart from white dwarfs and slowly pulsating B (SPB) stars.
For the hybrid $\gamma$ Dor--$\delta$ Sct stars, a comparison between observation and theory was
done by Bouabid et al.(2009). Using non-adiabatic computations, they predict the excitation
of both $\gamma$ Dor and $\delta$ Sct modes separated by a region of stable modes in models
located in the region of the H-R diagram where hybrid candidates have been detected.
The hybrid stars occupy the same area with $\delta$ Sct stars and is not confined to
the overlapping region in the H-R diagram.
They behave differently in the $g$-mode regime from $\gamma$ Dor stars.
They likely form a different class of pulsating stars and are not merely a simple mixture of
$\delta$ Sct and $\gamma$ Dor types pulsations.
Thus a different driving mechanism is demanded to excite the observed $g$-modes of hybrid stars.
A detailed discussion on $\delta$ Sct--$\gamma$ Dor hybrids can be found in
Grigahc\`{e}ne et al.(2010) and Hareter (2013).

  \item  Modified modelling is demanded to explain the existence of hybrid pulsations.
  Attempts towards seismic modelling of the observed hybrid pulsation frequencies,
  for the $\beta$ Cep--SPB, $\delta$ Sct--$\gamma$ Dor, and sdBV$_{\rm rs}$ are overwhelming.
\end{enumerate}

\section{Ending Remarks}
\label{sect:Remarks}
The goal of this survey is to have an overall view of various hybrid pulsators,
actually the variables with multiple variability types.
We have extracted the hybrid information based on the \href{http://simbad.u-strasbg.fr/simbad/}{\vp Simbad} astronomical database.
This report presents our preliminary results, the cross table Table~\ref{Tab:hybridPul}
along with several related statistics (Tables~\ref{Tab:BinPul}--\ref{Tab:percent}).
A web page \href{http://www.chjaa.org/COB/hybrid/}{\vp (http://www.chjaa.org/COB/hybrid/)} that allows one to retrieve
the hybrid stars of any two types of variability is developed.
We draw the readers' attention to several kinds of hybrid pulsators, including
pulsating stars in binary systems, pulsating stars hosting transiting exoplanets, and so on.
In the next step, we plan to update the table by checking literature and
other database. In the mean time, we are going to do observations on
some of the interesting hybrid pulsators.

\bigskip\bigskip
\begin{acknowledgements}
This research has made use of the SIMBAD database, operated at CDS, Strasbourg, France.
The author has consulted the GCVS for definitions and classifications of variable stars.
This research was funded by the National Natural Science Foundation of China (NSFC)
under the project No.~11373037.
\end{acknowledgements}

\clearpage
\thispagestyle{empty}
\hoffset 0.95 truecm
\voffset -0.335 truecm   
\setcounter{table}{1}
\begin{sidewaystable}
\begin{center}
\vspace{-10mm}
\begin{minipage}[c]{130mm}
\caption[]{ A Statistics on the Multiple Identities Hybrid Pulsating Variable Stars$^{\dag}$.
\label{Tab:hybridPul}
}  
\end{minipage}
 \begin{tabular}{lrrrrrrrrrrrrrrrrrrrr}
  \hline\noalign{\smallskip}
            &BCEP & DCEP& CEP & CV~~ &DSCT & GDOR& M~~~ & roAp& RR~~ & RV~~ & sdBV& SR  &SXPHE& WR~~ & CW~~ &ZZ~~ &WD~~ &HB~~ &BS~~ \\
            & bC* & cC* & Ce* & CV* & dS* & gD* & Mi* & a2* & RR* & RV* &sdB~~~ & sr* & SX*   & WR* & WV* & ZZ* & WD* & HB* & BS* \\
  \hline\noalign{\smallskip}
bC*         & 332 &   0 &   1 &   0 &   1 &   0 &   0 &   0 &   0 &  0  &   5 &   0 &   0 &   1 &   0 &   0 &  1  &   6 &   0 \\
cC*         &   0 &4432 &1970 &   2 &   2 &   1 &   0 &   0 &  42 &  2  &   0 &   2 &  18 &   0 &   6 &   0 &  1  &   1 &   0 \\
Ce*         &   1 &1970 &13710&   2 &   1 &   0 &   2 &   0 &  31 & 17  &   1 &   5 &   0 &   0 & 123 &   0 &  0  &   0 &   0 \\
CV*         &   0 &   2 &   2 &1060 &   1 &   0 &   1 &   0 &   0 &  0  &   1 &   2 &   0 &   0 &   0 &   5 & 29  &   0 &   0 \\
dS*         &   1 &   2 &   1 &   1 &5197 &   2 &   0 &   0 &  46 &  0  &   0 &   0 &   9 &   0 &   0 &   0 &  1  &  14 &   1 \\
gD*         &   0 &   1 &   0 &   0 &   2 & 717 &   0 &   2 &   1 &  0  &   0 &   0 &   0 &   0 &   0 &   0 &  0  &   0 &   1 \\
Mi*         &   0 &   0 &   2 &   1 &   0 &   0 &10837&   0 &   3 &  0  &   0 & 112 &   0 &   0 &   0 &   0 &  1  &   0 &   0 \\
a2*         &   0 &   0 &   0 &   0 &   0 &   2 &   0 & 515 &   0 &  0  &   0 &   0 &   0 &   0 &   0 &   0 &  0  &   0 &   0 \\
RR*         &   0 &  42 &  31 &   0 &  46 &   1 &   3 &   0 &62319&  1  &   0 &  14 &  39 &   0 &  22 &   0 &  0  & 144 &  16 \\
RV*         &   0 &   2 &  17 &   0 &   0 &   0 &   0 &   0 &   1 &262  &   0 &  13 &   0 &   0 &   8 &   0 &  0  &   0 &   0 \\
sdBV        &   5 &   0 &   1 &   1 &   0 &   0 &   0 &   0 &   0 &  0  &  57 &   0 &   0 &   0 &   0 &   1 &  3  &  13 &   0 \\
sr*         &   0 &   2 &   5 &   2 &   0 &   0 & 112 &   0 &  14 & 13  &   0 &20110&   0 &   1 &   2 &   0 &  0  &   0 &   0 \\
SX*         &   0 &  18 &   0 &   0 &   9 &   0 &   0 &   0 &  39 &  0  &   0 &   0 & 447 &   0 &   0 &   0 &  1  &   1 &  22 \\
WR*         &   1 &   0 &   0 &   0 &   0 &   0 &   0 &   0 &   0 &  0  &   0 &   1 &   0 & 1407&   0 &   0 &  0  &   0 &   0 \\
WV*         &   0 &   6 & 123 &   0 &   0 &   0 &   0 &   0 &  22 &  8  &   0 &   2 &   0 &   0 & 749 &   0 &  0  &   0 &   0 \\
ZZ*         &   0 &   0 &   0 &   5 &   0 &   0 &   0 &   0 &   0 &  0  &   1 &   0 &   0 &   0 &   0 & 220 &185  &   2 &   0 \\
AB*         &   0 &   0 &   2 &   0 &   1 &   0 &   0 &   0 &   0 &  4  &   0 & 699 &   0 &   0 &   0 &   0 &  0  &   0 &   0 \\
pA*         &   0 &   0 &   6 &   0 &   0 &   0 &   2 &   0 &   1 & 94  &   0 &  39 &   0 &   1 &   1 &   1 &  3  &   0 &   0 \\
RC*         &   0 &   0 &   0 &   0 &   0 &   0 &   6 &   0 &   0 &  0  &   0 &  25 &   0 &   0 &   0 &   0 &  0  &   0 &   0 \\
s*b         &   0 &   0 &   0 &   0 &   0 &   0 &   0 &   0 &   0 &  0  &   0 &   0 &   0 &  10 &   0 &     &  0  &   0 &   0 \\
Pu*         &  12 &   4 &   7 &   2 &  46 &  19 &   5 &   2 &   7 & 13  &  57 & 142 &   2 &   0 &  21 &   4 &  5  &  14 &   0 \\
blu         &   3 &   0 &   1 &  17 &   8 &   0 &   0 &   7 &  11 &  0  &   2 &   1 &   0 &   0 &   0 &  16 &927  & 134 &  10 \\
BS*         &   0 &   0 &   0 &   0 &   1 &   1 &   0 &   0 &  16 &  0  &   0 &   0 &  22 &   0 &   0 &   0 & 48  & 351 &4024 \\
HB*         &   6 &   1 &   0 &   0 &  14 &   0 &   0 &   0 & 144 &  0  &  13 &   0 &   1 &   0 &   0 &   2 & 61  &18633& 351 \\
sdB         &  10 &   0 &   1 &   1 &   0 &   0 &   0 &   0 &   0 &  0  &1120 &   0 &   0 &   0 &   0 &   1 &161  &  49 &   0 \\
WD*         &   1 &   1 &   0 &  29 &   1 &   0 &   1 &   0 &   0 &  0  &   3 &   0 &   1 &   0 &   0 & 185 &20567&  61 &  48 \\
El*         &   3 &   4 &   8 &   0 &  13 &   9 &   0 &   1 &   0 &  1  &   1 &   1 &   0 &   3 &   2 &   0 &  0  &   1 &   0 \\
Al*         &   0 &   2 &   1 &  12 &   4 &   8 &   0 &   0 &   9 &  0  &   0 &   3 &   1 &   5 &   1 &   0 &  5  &   0 &   0 \\
bL*         &   0 &   1 &   4 &   1 &   3 &   1 &   0 &   0 &   1 &  0  &   1 &   0 &   0 &   3 &   0 &   0 &  0  &   0 &   0 \\
BY*         &   0 &   0 &   0 &   1 &   0 &   1 &   0 &   0 &   0 &  0  &   0 &   1 &   0 &   0 &   0 &   0 &  2  &   0 &   0 \\
EB*         &   2 &  16 &  43 &   4 &  22 &  23 &   2 &   0 &  35 &  7  &   4 &  12 &   4 &   4 &  13 &   0 & 20  &   2 &   7 \\
SB*         &  14 &  42 &  21 &  24 &  36 &  13 &   1 &  35 &  26 &  0  &   3 &  26 &   3 &  47 &   2 &   0 &1007 &  51 &   4 \\
EP*         &   0 &   0 &   0 &   0 &   0 &   0 &   0 &   0 &   0 &  0  &   0 &   0 &   0 &   0 &   0 &   0 &  0  &   0 &   0 \\
WU*         &   3 &   2 &   5 &   1 &  12 &   3 &   0 &   0 &  43 &  0  &   0 &   1 &   0 &   0 &   0 &   0 &  1  &   1 &   2 \\
RS*         &   0 &  20 &  19 &   0 &   0 &   0 &   0 &   0 &   1 &  0  &   0 &   2 &   0 &   0 &   1 &   0 &  3  &   0 &   0 \\
  \hline\noalign{\smallskip}
\end{tabular}
\end{center}
\vspace{-3mm}\hspace{-50mm}$^{\dag}${\it Note: }Check the latest updates via the web version at~\url{http://www.chjaa.org/COB/hybrid/} .\\
\vspace{0.5mm}\centering {\bf page 7}
\end{sidewaystable}

\clearpage
\thispagestyle{empty}
\hoffset 0.75 truecm
\voffset 0.55 truecm   
\begin{sidewaystable}
\begin{center}
\vspace{-10mm}
\begin{minipage}[l]{160mm}
\caption[]{Multiple Identities Hybrids: the Intrinsic Pulsating Variable Stars in Binary Systems and others.
\label{Tab:BinPul}} 
\end{minipage}
 \begin{tabular}{lrrrrrrrrrrrrr}
  \hline\noalign{\smallskip}
          & EA  & BCEP& BY~~ & EP~~ & RS~~ & EW~~ & Be~~  &     &     & ELL & Em~~ & UV  &post-AGB\\
          & Al* & bL* & BY* & EP* & RS* & WU* & Be* & pr* &blu  & El* & Em* & UV  & pA* \\
  \hline\noalign{\smallskip}
bC*       &   0 &  0  &  0  &  0  &   0 &  3  & 12  &  0  &  3  &  3  & 20  & 174 &  0  \\
cC*       &   2 &  1  &  0  &  0  &  20 &  2  &  0  &  4  &  0  &  4  &  0  &  21 &  0  \\
Ce*       &   1 &  4  &  0  &  0  &  19 &  6  &  0  &  2  &  1  &  8  & 19  &  19 &  6  \\
CV*       &  12 &  1  &  1  &  0  &   0 &  1  &  0  &  0  & 17  &  0  & 13  &  81 &  0  \\
dS*       &   4 &  3  &  0  &  0  &   0 & 12  &  0  &  1  &  8  & 13  &  3  & 310 &  0  \\
gD*       &   8 &  1  &  1  &  0  &   0 &  3  &  0  &  0  &  0  &  9  &  2  &  72 &  0  \\
Mi*       &   0 &  0  &  0  &  0  &   0 &  0  &  0  &  1  &  0  &  0  &188  &   1 &  2  \\
a2*       &   0 &  0  &  0  &  0  &   0 &  0  &  0  &  0  &  7  &  1  &  1  & 357 &  0  \\
RR*       &   9 &  1  &  0  &  0  &   1 & 43  &  0  &  2  & 11  &  0  &  5  &  49 &  1  \\
RV*       &   0 &  0  &  0  &  0  &   0 &  0  &  0  &  1  &  0  &  1  & 10  &   1 & 94  \\
sdBV      &   5 &  1  &  0  &  0  &   0 &  0  &  0  &  0  &  2  &  1  &  0  &  15 &  0  \\
sr*       &   3 &  0  &  1  &  0  &   2 &  1  &  0  &  2  &  1  &  1  &120  &  21 & 39  \\
SX*       &   1 &  0  &  0  &  0  &   0 &  0  &  0  &  0  &  0  &  0  &  0  &   1 &  0  \\
WR*       &   5 &  3  &  0  &  0  &   0 &  0  &  2  &  0  &  0  &  3  &261  &  75 &  1  \\
WV*       &   1 &  0  &  0  &  0  &   1 &  0  &  0  &  0  &  0  &  2  &  0  &   0 &  1  \\
ZZ*       &   0 &  0  &  0  &  0  &   0 &  0  &  0  &  0  & 16  &  0  &  4  &  68 &  1  \\
AB*       &   0 &  0  &  0  &  0  &   0 &  0  &  0  &  0  &  0  &  0  &  8  &   1 & 12  \\
Al*       &6634 & 72  &  1  &  0  &  10 & 37  &  2  & 39  &  7  &  2  & 48  & 377 &  0  \\
bL*       &  72 &1590 &  0  &  0  &   1 & 46  &  6  & 19  &  0  &  3  & 36  & 199 &  1  \\
blu       &   7 &  0  &  0  &  0  &   1 &  0  &  7  &  0  &19335&  2  & 11  &2019 &  9  \\
BS*       &   0 &  0  &  0  &  0  &   0 &  2  &  0  &  0  & 10  &  0  &  0  &   7 &  0  \\
BY*       &   1 &  0  &978  &  0  &   3 &  0  &  0  & 63  &  0  &  0  & 40  & 104 &  0  \\
EB*       &1189 &193  &  3  &  1  &   5 &618  &  5  & 22  &  0  &113  & 36  & 141 &  0  \\
El*       &   2 &  3  &  0  &  0  &   1 &  8  &  3  &  0  &  2  &501  & 13  &  95 &  0  \\
EP*       &   0 &  0  &  0  & 31  &   0 &  0  &  0  &  0  &  0  &  0  &  0  &   0 &  0  \\
HB*       &   0 &  0  &  0  &  0  &   0 &  1  &  0  &  0  &134  &  1  &  0  &  90 &  0  \\
pA*       &   0 &  1  &  0  &  0  &   0 &  1  &  3  &  1  &  9  &  0  & 39  &  24 &454  \\
Pu*       &   5 &  3  &  2  &  0  &   1 &  7  & 19  &  4  &  5  & 30  &126  & 299 & 22  \\
RC*       &   1 &  0  &  0  &  0  &   0 &  0  &  0  &  0  &  0  &  0  &  2  &   1 & 35  \\
RS*       &  10 &  1  &  3  &  0  & 511 &  5  &  0  &  8  &  1  &  1  & 10  &  94 &  0  \\
s*b       &   1 &  4  &  0  &  0  &   0 &  0  &  0  &  0  &  3  &  3  & 78  & 155 &  1  \\
SB*       & 329 &136  & 35  &  2  & 184 &153  & 31  & 43  &131  & 68  &232  &1468 &  4  \\
sdB       &   5 &  1  &  0  &  0  &   0 &  0  &  0  &  0  &100  &  2  &  0  & 398 &  0  \\
WD*       &   5 &  0  &  2  &  0  &   3 &  1  &  0  &  1  &927  &  0  &  9  &1834 &  3  \\
WU*       &  37 & 46  &  0  &  0  &   5 &5064 &  0  & 24  &  0  &  8  &  1  &  43 &  1  \\
  \hline\noalign{\smallskip}
\end{tabular}
\end{center}
\vspace{0.5mm}\centering {\bf page 8}
\end{sidewaystable}

\clearpage
\begin{table}[th!]
\caption[c]{A Statistics on the Major Types of Stars Archived by Simbad.
Third and Fourth Columns Refer to the Ratios to All Variables and to All Stars, Respectively. }
   \label{Tab:percent}
   \vspace{-5mm}
   \begin{center}
   \begin{tabular}{lrrcl}
\hline\noalign{\smallskip}
 \multicolumn{1}{c}{Type of Pulsators}&  \multicolumn{1}{c}{Numbers}  &  \multicolumn{1}{r}{Percentage} &   \multicolumn{1}{c}{Ten-Thousandth}&\multicolumn{1}{l}{Type of Pulsators}     \\
 (Simbad's ID) &   & \multicolumn{1}{r}{ (1/100)} & \multicolumn{1}{c}{ (1/10000)} &(Generic Names) \\
  \hline\noalign{\smallskip}
RR*         &62319 &25.12    &165.96   & RR Lyr-type stars        \\
WD*         &20567 & 8.29    &~54.77   & White Dwarfs             \\
sr*         &20110 & 8.11    &~53.55   & semiregular variables    \\
blu         &19335 & 7.79    &~51.49   & blue objects             \\
EB*         &19234 & 7.75    &~51.22   & Eclipsing Binaries       \\
HB*         &18633 & 7.51    &~49.62   & Horizontal Branch stars  \\
Ce*         &13710 & 5.52    &~36.51   & Cepheids                 \\
Mi*         &10837 & 4.36    &~28.86   & Mira-type                \\
Pu*         & 7738 & 3.12    &~20.61   & Pulsating variable stars \\
SB*         & 6985 & 2.81    &~18.60   & Spectroscopic Binaries   \\
Al*         & 6634 & 2.67    &~17.67   & Algol-type EB            \\
dS*         & 5197 & 2.09    &~13.84   & $\delta$ Scuti stars     \\
WU*         & 5064 & 2.04    &~13.48   & W UMa-type EB            \\
AB*         & 4435 & 1.79    &~11.81   & Asymptotic Giant Branch stars                      \\
cC*         & 4432 & 1.79    &~11.80   & classical Cepheids       \\
BS*         & 4024 & 1.62    &~10.71   & Blue Straggler stars     \\
bL*         & 1590 & 0.64    &~~4.23   & $\beta$ Lyr-type\\
WR*         & 1407 & 0.57    &~~3.75   & Wolf-Rayet stars\\
sdB         & 1120 & 0.45    &~~2.98   & subdwarf B stars\\
CV*         & 1060 & 0.43    &~~2.82   & Cataclysmic Variables\\
BY*         &  978 & 0.39    &~~2.60   & BY Dra-type\\
WV*         &  749 & 0.30    &~~1.99   & W Vir-type\\
gD*         &  717 & 0.28    &~~1.90   & $\gamma$ Dor stars       \\
s*b         &  542 & 0.22    &~~1.44   & luminous blue variables  \\
a2*         &  515 & 0.21    &~~1.37   & $\alpha^2$ CVn, roAp     \\
RS*         &  511 & 0.20    &~~1.36   & RS CVn-type              \\
El*         &  501 & 0.20    &~~1.33   & ellipsoidal variables    \\
pA*         &  453 & 0.18    &~~1.20   & post AGB                 \\
SX*         &  447 & 0.18    &~~1.19   & SX Phe-type              \\
bC*         & 332  & 0.13    &~~0.88   & $\beta$ Cep-type         \\
RV*         &  262 & 0.11    &~~0.69   & RV Tau-type              \\
ZZ*         &  220 & 0.09    &~~0.58   & ZZ Cet type              \\
RC*         &  150 & 0.06    &~~0.39   & R CrB-type               \\
sdBV        &   57 & 0.02    &~~0.15   & pulsating sdB            \\
EP*         &   31 & 0.01    &~~0.08   & eclipses by planet       \\
***         &97421 & 2.59    &  \multicolumn{2}{l}{~~~~~~~~double/multiple stars vs all stars} \\
V*          &248103& 6.61    &  \multicolumn{2}{l}{~~~~~~~~all variables in stars as of Dec.31, 2014}\\
**          &3754965&---     &  \multicolumn{2}{l}{~~~~~~~~all stars as of Dec.31, 2014}\\
GCVS4.2     & 58947&23.7     &  \multicolumn{2}{l}{~~~~~~~~GCVS' total variables vs the Simbad's} \\
   \hline\noalign{\smallskip}
   \end{tabular}
   \end{center}
\end{table}

\clearpage

\voffset = 6 mm
\hoffset = 2 mm
\begin{longtable}{lrl}
\vspace{-0.20mm}\\
\caption{ Selected Interesting Hybrid Pulsators with Dual or Multiple Identities.
The Reality of Multiple Identities Are Not Examined. 
Check the Entries at: \url{http://www.chjaa.org/COB/hybrid/.}}
\label{Tab:demo-hybrid}
\vspace{-1.0mm}\\
\hline\noalign{\smallskip}
Hybrid Identities$^{\dag}$& N &  \multicolumn{1}{c}{Notes}  \\
\hline\noalign{\smallskip}
\endfirsthead  

\multicolumn{3}{c} {{ \tablename\ \thetable{} -- continued from previous page}}
\vspace{3.0mm} \\
\hline\noalign{\smallskip}
Hybrid Identities$^{\dag}$& N  & \multicolumn{1}{c}{Notes}  \\
\hline\noalign{\smallskip}
\endhead   

\hline\noalign{\smallskip}
\vspace{-3.0mm}\\
\multicolumn{3}{c}{{ Continued on next page}} \\
\endfoot  

\vspace{-3.0mm}\\
\multicolumn{3}{c}{{The end of Table 5}} \\
\endlastfoot  

a2+Em+UV          &   1 &~~V1045 Ori=HD 36916          \\
AB+pA+UV          &   1 &~~NGC 5904 MSB 1, AB=Asymptotic Giant Branch Star (AGB)   \\
Al+SX             &   1 &~~KIC 10989032: SX Phe-type star in Algol-type EB\\
bC+Ce+UV          &   1 &~~V757 Per    \\
bC+El+HB+Pu+sdB   &   1 &~~2MASS J19343993+4758117      \\
bC+Em+WR          &   1 &~~V1035 Sco                    \\
bC+HB+El          &   1 &~~2MASS J19343993+4758117  \\
bC+UV+WD          &   1 &~~lam Sco  \\
bL+CV             &   1 &~~KIC 5112741    \\
bL+pA             &   1 &~~HP Lyr, EB of $\beta$ Lyr-type, semi-detached\\
BS+gD             &   1 &~~KIC 5024455 \\
BY+CV             &   1 &~~NGC104 EGG V27         \\
cC+**             &  30 &~~cC in double/multiple stellar systems\\
cC+bL+Ce          &   1 &~~V480 Lyr \\
cC+Ce             &1970 &~~ cC=classical Cepheids=$\delta$ Cep type (DCEP) \\
cC+Ce+RR          &   9 &~~ Ce=Cepheids, RR=RR Lyr-type \\
Ce+Al             &   1 &~~KIC 2447893, EA+Ce\\
Ce+blu            &   1 &~~TYC 656-628-1   \\
Ce+RR             &  31 &~~ Ce*=Cepheids \\
CV+AM+sr+PM+blu+UV&   1 &~~AR UMa: (CV of AM Her type, polar) \\
dS+AB             &   1 &~~V5505 Sgr, $\delta$ Sct star\\
dS+bC             &   1 &~~V1228 Cen=HD 100495\\
dS+blu+UV         &   2 &~~ER Leo, DT CVn   \\
dS+BS+blu         &   1 &~~PHL 319 \\
dS+BS+HB          &   1 &~~SDSS J022422.40+004500.8   \\
dS+Ce             &   1 &~~KIC 9594189: \\
dS+CV             &   1 &~~V1209 Tau        \\
dS+pr             &   1 &~~V1247 Ori \\
dS+RR             &  46 &~~ 1 SB, e.g. UY Cam, SZ Lyn, AD CMi,VZ Cnc,BS Aqr $\cdots$ \\
dS+RR+UV          &   4 &~~UV=Eruptive variables of the UV Ceti type\\
dS+SX             &   9 &~~ 1 EB; BQ Psc: SX+HB+dS+cC\\
dS+WD+SB+UV       &   1 &~~IK Peg   \\
El+a2+SB+UV       &   1 &~~ 33 Tau (=V817 Tau =HD 24769)\\
El+RS+UV+SB       &   1 &~~$\iota$ Tri\\
El+SB+Pu          &   1 &~~ HY Vel                      \\
gD+bL             &   1 &~~2MASS J18441225+0613529\\
gD+BY             &   1 &~~FG CVn     \\
gD+cC             &   1 &~~2MASS J06522020-0535137 \\
Mi+CV             &   1 &~~FQ Mon          \\
Mi+pr             &   1 &~~HV 1644,  Mira Cet type\\
Mi+WD+PM+Em+UV+Mas&   1 &~~omi Cet,  Mira Cet type\\
pA+Em+WR          &   1 &~~OH 284.2 -0.8, post-AGB star \\
Pu+SB+CV          &   1 &~~4 Dra            \\
RC+Al             &   1 &~~V532 Oph, R CrB type\\
RC+pA+Em+UV       &   1 &~~R CrB, R CrB type\\
RR+bL             &   1 &~~  TZ Cap  \\
RR+gD             &   1 &~~2MASS J06520507-0511481=CoRoT 110679591\\
RR+SX             &  39 &~~e.g. XX Cyg, AE UMa, CY Aqr, DY Peg $\cdots\cdots$  \\
RR+SX+cC          &   7 &~~e.g. XX Cyg, V1638 Oph, V879 Her $\cdots\cdots$  \\
RS+bL+Em+pr       &   1 &~~FI Cru   \\
RS+blu+SB         &   1 &~~EZ Peg   \\
RS+Pu+SB          &   1 &~~BM Lyn                 \\
RS+RR+SB          &   1 &~~AG Cnc      \\
RS+SB+blu         &   1 &~~EZ Peg    \\
RV+El+Ce          &   1 &~~HV 1369 \\
RV+pA+UV          &   1 &~~AC Her    \\
RV+RR             &   1 &~~V1127 Aql  \\
sdb+Pu+ZZ+HB      &   1 &~~2MASS J19090714+3756143      \\
sdBV+EB+bL+Ce     &   1 &~~TYC 3556-3568-1  \\
sdBV+EB+CV+Pu     &   1 &~~NAME KBS 13      \\
sr+BY             &   1 &~~GQ Vel     \\
sr+El             &   1 &~~V5485 Sgr: semi-regular pulsating star\\
sr+pr+RV+pA       &   1 &~~UY CMa: semi-regular pulsating star\\
sr+WR+pA?+Mas     &   1 &~~PZ Cas: double stars\\
sr+WU             &   1 &~~V2616 Oph  \\
SX+dS+HB+cC       &   1 &~~BQ Psc  \\
SX+WD+cC          &   1 &~~BL Cam   \\
WD+pr+UV+SB       &   1 &~~HD 217411\\
WU+bL+dS          &   1 &~~FH Cam  \\
WU+CV             &   1 &~~NGC 104 EGG V7  \\
WU+dS+bL+UV       &   1 &~~CC Lyn           \\
WU+Em+pr          &   1 &~~V1537 Ori        \\
WU+HB             &   1 &~~BD Gru\\
WU+pA+RR          &   1 &~~V1017 Cyg, EB of W UMa type, contact\\
WU+WD+UV          &   1 &~~NGC 104 EGG V20     \\
WV+Al             &   1 &~~V538 Ara   \\
WV+pA             &   1 &~~ST Pup   \\
WV+RR+Ce          &   3 &~~WV=variables of W Vir type \\
WV+RS             &   1 &~~MX Dra \\     
ZZ+pA             &   1 &~~LW Lib,  Pulsating White Dwarf   \\
   \hline\noalign{\smallskip}
\end{longtable}

\clearpage
\hoffset = -5 mm

\begin{table*}[th!]
\begin{minipage}[c]{160mm}
\caption[c]{Lookup Table of the Surveyed Pulsators: the Simbad Object Types against GCVS Acronyms.
   \label{Tab:object-type} } 
\end{minipage}
   \vspace{-5mm}
   \begin{center}
   \begin{tabular}{llrl}
\hline\noalign{\smallskip}
Simbad & GCVS     & Numbers  & \multicolumn{1}{c}{Extended explanation}   \\
   \hline\noalign{\smallskip}
\multicolumn{3}{l}{
1.~ Pulsating Variables}\\
\cline{1-2}
---   &ACYG       &  64 & $\alpha$ Cygni-type non-radial pulsating supergiant stars of spectral types B or A \\
bC*   &BCEP       &  332& Pulsating Variable Star of $\beta$ Cephei-type, multiperiodic  \\
bL*   &EB(b)      & 1590& Eclipsing binary of $\beta$ Lyr type (semi-detached) \\
BS*   &BSS        & 4024& Blue Straggler Stars:  main-sequence stars in open/globular clusters that are \\
      &           &     & more luminous and bluer than stars at the MS turn-off point for the cluster. \\
cC*   &DCEP       & 4432& classical Cepheids (i.e. $\delta$ Cephei-type variables, pop. I, luminous, massive)\\
Ce*   &CEP        &13710& Cepheids, variable stars of radially pulsating, high luminosity  \\
DQ*   &---        &   40& CV DQ Her type (intermediate polar), e.g. FO Aqr, XY Ari, V667 Pup, PQ Gem\\
dS*   &DSCT       & 5197& Pulsating Variable Star of $\delta$ Scuti-type, radial,nonradial, pressure modes\\
gD*   &GDOR       &  717& Pulsating Variable Star of $\gamma$ Doradus-type, gravity modes        \\
HB*   &HB         &18633& Horizontal Branch (HB) stars: at a stage of stellar evolution that immediately follows RGB\\
sdB   &sdB/sdO/RPHS&  57& sdBV/sdOV: pulsating subdwarf B/O variable stars: very hot and bright subdwarf stars \\
      &           &     & (core helium buring, $T_{\rm eff}\sim$ 20,000--40,000\,K, $\log g\sim$5.0--6.2, M$\sim$0.5\,$M_{\odot}$) with spectral\\
      &           &     &   type B or O, situated at the extreme/hottest horizontal branch of the H-R diagram unlike\\
      &           &     &  normal horizontal branch stars (Herber 2008, 1986). Three sub-types:  \\
      &           &     & (1) short-period(rapid) $p$-mode: sdBVr: P=90--600s, prototype EC 14026=V361 Hya:\\
      &           &     &      PG 1605+072, PG 1047+003, PG 0014+067; \\
      &           &     & (2) long-period(slow) $g$-mode: sdBVs (or LPsdBV), P=45--180m, PG 1716+426=V1093 Her; \\
      &           &     & (3) hybrid $p$- and $g$-mode: sdBVrs, DW Lyn=HS 0702+6043.\\
blu   &---        &19335& blue (horizontal branch) objects\\
LPB   &LPB        &  ---& Long-period pulsating B stars (P$>$1\,d)\\
Mi*   &M          &10837& Mira (Omicron Cet): Variable Star of Mira Cet type\\
---   &PVTEL      &  ---& PV Telescopii-type pulsators\\
Pu*   &---        & 7738& Pulsating variable Star\\
RR*   &RR         &62319& RRab/RRc/RRd: RR Lyr variables: radially-pulsating, old Pop.II giant in globular clusters\\
RV*   &RV         &  262& type-II Cepheids of RV Tauri-type: radially pulsating yellow supergiants, \\  
WV*   &CW         &  749& CWA, CWB: type-II Cepheids of W Virginis-type, BL Her-type\\
sr*   &SR         &20110& Semiregular pulsating variables: sub-types: SRA,SRB,SRC,SRD \\ 
SX*   &SXPHE      &  447& SX Phe-type variables: pulsating subdwarfs, old disk galactic Pop. in globular clusters \\
ZZ*   &DAV,DBV,DOV&  220& pulsating white dwarfs of ZZ Ceti type: ZZA, ZZB      \\
---   &SPB        &  ---& slowly pulsating B star(SPB), prototype 53 Persei: main-sequence stars of \\
      &           &     & spectral types B2 to B9, 6 had been found to exhibit both SPB+ beta Cephei variability.\\
a2*   &roAp       &  515& Variable Stars of Alpha$^{2}$ CVn-type (or $\alpha^2$ CVn variable): chemically peculiar\\
      &           &     &  main-sequence stars of spectral class B8p to A7p. They have strong magnetic fields and     \\
      &           &     &  strong silicon, strontium, or chromium spectral lines. The chemically peculiar roAp       \\
      &           &     &  (rapidly oscillating A-type peculiar stars, or Przybylski's stars) occupy the MS end \\
      &           &     &  of the $\delta$ Sct instability strip. \\
   \hline\noalign{\smallskip}
   \end{tabular}
   \end{center}
\end{table*}

\begin{table*}[th]
\addtocounter{table}{-1}
\caption[c]{ --- Continued}
   \vspace{-5mm}
   \begin{center}
   \begin{tabular}{llrl}
\hline\noalign{\smallskip}
Simbad & GCVS     & Numbers  & Extended explanation   \\
   \hline\noalign{\smallskip}

\multicolumn{3}{l}{
2. Eruptive variables:}\\
\cline{1-2}
RS*   &RS         &  511& Eruptive variables of the RS CVn type: Close binaries with chromospheric activity, \\
      &           &     & causing very small light variations. Eclipses and X-ray variability often seen as well.\\
s*b   &LBV/SDOR   &  542& Luminous Blue Variables (LBVs): massive evolved, very luminous blue supergiant stars, \\
      &           &     & usually surrounded by expanding envelopes known as S Doradus-type variables: \\
      &           &     & unpredictable, dramatic variations in both spectra and brightness, often associated with nebulae. \\
      &           &     & Occasional outbursts up to 7 magnitudes, lasting for months, caused by ejection of a shell of \\
      &           &     & matter, e.g. P Cyg, $\eta$ Carinae, AG Car, AE And, AF And. \\
UV    &UV         &91387& Eruptive variables of the UV Ceti type, UV-emission source, red dwarf stars showing outbursts \\
      &           &     & up to 6 magnitudes lasting for only a few minutes, caused by flares.\\
WR*   &WR         & 1407& Systems having Eruptive Wolf-Rayet stars. Emission lines of carbon \& nitrogen and evidence \\
      &           &     & for unstable mass outflow as a ``stellar wind".\\
RC*   &RCB        &  150& pulsating variables of the R CrB type: hydrogen-poor, carbon- and helium-rich, \\
      &           &     & high-luminosity stars, simultaneously eruptive and pulsating variables\\
Em*   &Em         &21353& Emission-line stars \\
Er*   &Er         &   68& Eruptive variable Star\\
FU*   &FU         &   40& Variable Star of FU Ori type\\
LP*   &LPV        &76895& Long-period variable star, e.g. FZ Cas, V475 Cyg, CI Cas.   \\
\multicolumn{3}{l}{
3. Eclipsing Binary Systems         } \\
\cline{1-3}
Al*   &EA         & 6634& Algol-type eclipsing binary systems (detached), EB of Algol ($\beta$ Persei), \\
EB*   &E          &19234& Eclipsing binary systems\\
bL*   &EB(b)      & 1590& $\beta$ Lyrae-type eclipsing binary systems, having ellipsoidal components\\
WU*   &EW         & 5064& W Ursae Majoris-type eclipsing binaries (contact).\\
EP*   &---        &   31& Star showing eclipses by its planet -- planetary transit\\
RS*   &RS         &  511& Variable of RS CVn type, close binary, non-pulsation, starspots-caused variability\\
SB*   &SB         & 6985& Spectroscopic binary\\
WD*   &WD         &20567& WDA,WDB: Systems with White-Dwarf components   \\
XB*   &XB         & 1422& close Binary systems showing X-ray and optical bursts.  \\
\multicolumn{3}{l}{
4. Rotating variables}\\
\cline{1-2}
BY*   &BY         &  978& BY Draconis-type variables: non-pulsation, multiple stars, starspots-caused variability,\\
      &           &     & many are UV variables \\
El*   &ELL        &  501& Rotating ellipsoidal variables: elliptical, tidally distorted, close binary stars \\
      &           &     & with ellipsoidal components, no eclipses \\
\multicolumn{4}{l}{
5. Cataclysmic (Explosive and Novalike) Variables}\\
\cline{1-3}
AB*   &AGB        & 4435& Asymptotic Giant Branch stars: bright red giant (He-burning).\\
pA*   &---        &  453& post-AGB (Asmptotic Giant Branch) stars \\
RG*   &RGB        &15437& Red Giant Branch stars\\
AM*   &---        &   91& CV of AM Her type (polar), e.g. EP Dra, ES Cet, AN UMa, AR UMa, VV Pup\\
CV*   &CV         & 1060& Cataclysmic Variable stars      \\
DN*   &DN         &  648& Dwarf Novae\\
SN*   &SN         & 9066& SuperNovae\\
NL*   &NL         &  110& Nova-like variables\\
No*   &N          & 1788& Nova, close binary\\
Sy*   &---        &  231& Symbiotic Stars \\
\multicolumn{3}{l}{
6. non-pulsating variables}\\
\cline{1-3}
pr*   &           & 6105& Pre-main sequence stars\\
   \hline\noalign{\smallskip}
   \end{tabular}
   \end{center}
\end{table*}

\clearpage

\label{lastpage}


\begin{thebibliography}{999}

\bibitem{bar09} Baran, A., \& Fox-Machado, L. 2009, \apss, (arXiv:0912.4332)
\bibitem{bar05} Baran, A., Pigulski, A., Kozie\l, D., et al. 2005, MNRAS, 360, 737: Balloon 090100001
\bibitem{bar11} Baran, A., et al. 2011, MNRAS, 413, 2838:
\bibitem{bou09} Bouabid, M.-P., Montalb\'{a}n, J., Miglio, A., et al. 2009, in AIP Conf. Proc. Ser. 1170, Stellar Pulsation:
Challenges for Theory and Observation, ed. J. A. Guzik \& P. A. Bradley
(Melville, NY: AIP), 477 (arXiv:0911.0775):
                {\it Hybrid $\gamma$ Doradus/$\delta$ Scuti Stars: Comparison Between Observations and Theory}
\bibitem{bru14} Brunsden, E., Pollard, K. R., Cottrell, P. L., et al. 2014, MNRAS, accepted (arXiv:1412.2828): HD 49434
\bibitem{cha06} Chapellier E, Le Contel, D, Le Contel, J. M., et al. 2006, A\&A, 448, 697
\bibitem{cha13} Chapellier, E., \& Mathias, P. 2013, \aap, 556, 87\\
{\it The CoRoT star ID 100866999: a hybrid $\gamma$ Doradus--$\delta$ Scuti star in an eclipsing binary system}
\bibitem{cha12} Chapellier, E., Mathias, P., Weiss, W. W., et al. 2012, \aap, 540, 117\\
{\it Strong interactions between g- and p-modes in the hybrid $\gamma$ Dor--$\delta$ Sct CoRoT star ID 105733033}
\bibitem{das10} Daszy\'{n}ska-Daszkiewicz, J., \& Walczak, P. 2010, \apss, 328, 97 (arXiv:0908.4166): nu Eri
\bibitem{dup04} Dupret, M. A., et al. 2004, \aap, 414, L17
\bibitem{dup09} Dupret, M. A., Miglio, A., Montalb\'{a}n, J., et al. 2009, arXiv:0907.2636\\
{\it Driving and damping mechanisms in hybrid pressure-gravity modes pulsators}
\bibitem{dzi08} Dziembowski, W. A., \& Pamyatnykh, A. A. 2008, MNRAS, 385, 2061: nu Eri, 12 Lac
\bibitem{fon03} Fontaine, G., Brassard, P., Charpinet, S, et al. 2003, ApJ, 597, 518
\bibitem{gri10} Grigahc\`{e}ne, A.,  Antoci, V., Balona, L., et al. 2010, ApJL, 713, 192
\bibitem{han09b} Handler, G., 2009, MNRAS, 398, 1339: HD 8801, gamma Peg
\bibitem{han09} Handler, G., 2009, CoAst, 159, 42: {\it ``Hybrid'' pulsators -- fact or fiction?}
\bibitem{han12} Handler, G. 2012, ASPC, 462, 111 (arXiv:1112.5981):
{\it Hybrid Pulsators among A/F-type Stars}
\bibitem{han97} Handler, G., Kanaan, A., \& Montgomery, M. H. 1997, \aap, 326, 692: HS 2324+3944
\bibitem{han04} Handler, G, Shobbrook. R. R., Jerzykiewicz, M., et al. 2004, MNRAS, 347, 454
\bibitem{han06} Handler, G, Jerzykiewicz, M., Rodr\'{\i}guez, E, et al. 2006, MNRAS, 365, 327
\bibitem{har13} Hareter, M. M., 2013, PhD Thesis,
{\it Investigation of $\gamma$ Dor -- $\delta$ Sct hybrid stars based on high precision
space photometry and complementary ground based spectroscopy}, Universitat Wien.
\bibitem{har11} Hareter, M., Fossati, L., Weiss, W., et al. 2011, ApJ, 743, 153\\
{\it Looking for a Connection between the Am Phenomenon and Hybrid delta Sct -gamma Dor Pulsation:
Determination of the Fundamental Parameters and Abundances of HD 114839 and BD +18 4914}
\bibitem{heb86} Heber, U. 1986, \aap, 155, 33
\bibitem{heb08} Heber, U. 2008, Mem. S. A. It., 79, 375
\bibitem{hen05} Henry, G. W. \& Fekel, F. C. 2005, AJ, 129, 2026
\bibitem{her12} Herzberg, W., Uytterhoeven, K. \& Roth, M. 2012, AN, 333, 1077\\
{\it Ground-based multi-color photometry of the gamma Doradus--delta Scuti hybrid star KIC 6761539}
\bibitem{135} Jeffery, C. S. 2008, CoAst, 157, 240
\bibitem{kin06} King, H., Matthews, J. M., Row, J. F., et al. 2006, CoAst, 148, 28 (arXiv:0706.1804)\\
{\it HD 114839 - An Am star showing both delta Scuti and gamma Dor pulsations discovered through MOST photometry}
\bibitem{kur15} Kurtz, D. W., Hambleton, K. M., Shibahashi, H., et al. 2015, MNRAS, 446, 1223\\
{\it Validation of the frequency modulation technique applied to the pulsating delta Sct--gamma Dor eclipsing binary star KIC 8569819}
\bibitem{lec01} Le Contel, J.-M., Mathias, P., Chapellier, E., \& Valtier, J.-C. 2001, \aap, 380, 277: 53 Psc
\bibitem{lut08} Lutz, R., et al. 2008, in Hot Subdwarf Stars and Related Objects, eds. U. Heber, S. Jeffery, \& R. Napiwotzki, ASP Conf., 392, p.339
\bibitem{lut09} Lutz, R., Schuh, S., Silvotti, R., et al. 2009, \aap, 496, 469: V391 Pegasi
\bibitem{mac14} Maceroni, C., Lehmann, H., Da Silva, R., et al. 2014, \aap, 563, 59 (arXiv:1401.3130): KIC 3858884, hybrid DSCT\\
{\it KIC 3858884: a hybrid {delta} Scuti pulsator in a highly eccentric eclipsing binary.}
\bibitem{mig07} Miglio, A, Montalb\'{a}n., J., \& Dupret, M.-A. 2007, MNRAS, 375, L21
\bibitem{pam99} Pamyatnykh, A. A. 1999, Acta Astron. 49, 119
\bibitem{pan11} Pandey, C. P., et al. 2011, CoAst, 162, 21
\bibitem{reed10} Reed, M. D., Kawaler, S. D., {\O}stensen, R. H., et al. 2010, MNRAS, 409, 1496
\bibitem{rip11} Ripepi, V. et al. 2011, MNRAS, 416, 1535: CoRoT 102699796
\bibitem{row06} Rowe, J. F., Matthews, J. M., Cameron, C., et al. 2006, CoAst, 148, 34\\
{\it Discovery of hybrid gamma Dor and delta Sct pulsations in BD+18 4914 through MOST spacebased photometry}
\bibitem{sam07} Samus, N. N., Durlevich, O. V., Kazarovets, E. V., Kireeva, N. N., Pastukhova, E. N.,
Zharova, A. V., et al., {\it General Catalogue of Variable Stars}: \url{http://www.sai.msu.su/gcvs/gcvs/index.htm}
\bibitem{sch06} Schuh, S., Huber, J., Dreizler, S., et al. 2006, \aap, 445, L31.
\bibitem{sea05} Seaton, M. J. 2005 MNRAS, 362, L1
\bibitem{sil07} Silvotti, R., et al. 2007, Nature, 449, 189
 \bibitem{sod14} S\'{o}dor, \'{A.}, De Cat, P., Wright, D. J., et al. 2014, MNRAS, 438, 3535 (arXiv:1312.6307): HD 25558: SB2, SPB+SPB
\bibitem{tka13} Tkachenko, A., et al., 2013, \aap, 556, 52:
{\it Detection of a large sample of {gamma} Doradus stars from Kepler space photometry and high-resolution ground-based spectroscopy}
\bibitem{vau05} Vauclair, G., Solheim, J.-E., \& {\O}stensen, R. H. 2005, \aap, 433, 1097: Abell 43
\bibitem{zdr08} Zdravkov, T., \& Pamyatnykh A. A. 2008, CoAst, 157, 385 (arXiv:0810.1609)
\bibitem{zxb12} Zhang, X. B., Deng, L. C., Luo, C. Q. 2012, AJ, 144, 141:\\
{\it      A Probable Hybrid gamma Dor-delta Scuti Variable Discovered in the Open Cluster NGC 2126}
\bibitem{zay10} Zhou, A.-Y., 2010, \href{http://arxiv.org/abs/1002.2729}{arXiv:\vp1002.2729v6} (last revised 9 Dec 2014)
\end{thebibliography}
\end{document}